\documentclass[manuscript, screen, nonacm]{acmart}
\usepackage{threeparttable} 

\AtBeginDocument{%
  }

\begin{document}

\title{``Can You Play Anything Else?'' Understanding Play Style Flexibility in League of Legends}

\author{Emily Chen, Alexander Bisberg, Emilio Ferrara}

\affiliation{%
  \institution{University of Southern California}
  \city{Los Angeles}
  \state{California}
  \country{USA}
}

\renewcommand{\shortauthors}{Chen et al.}
\makeatletter
\let\@authorsaddresses\@empty
\makeatother
\begin{abstract}
This study investigates the concept of flexibility within League of Legends, a popular online multiplayer game, focusing on the relationship between user adaptability and team success. 
Utilizing a dataset encompassing players of varying skill levels and play styles, we calculate two measures of flexibility for each player: overall flexibility and temporal flexibility. 
Our findings suggest that the flexibility of a user is dependent upon a user's preferred play style, and flexibility does impact match outcome. This work also shows that skill level not only indicates how willing a player is to adapt their play style but also how their adaptability changes over time.
This paper highlights the duality and balance of specialization versus flexibility, providing insights that can inform strategic planning, collaboration and resource allocation in competitive environments.   
\end{abstract}

\maketitle

\section{Introduction}
In the dynamic ecosystem of gaming, where play styles evolve based on objectives, collaboration and competitiveness, player experiences are anything but static. This fluidity motivates the core of our investigation into League of Legends (LoL), a Multiplayer Online Battle Arena (MOBA) game published in 2009 by Riot Games. 

LoL's main game play centers around two teams of five tasked with destroying the opposition's base, underscoring the intricate balance between individual skill and team synergy needed for success \cite{kim2017makes, sapienza2018individual}. Unlike previous studies that have explored matchmaking \cite{claypool2015surrender}, user archetypes \cite{eggert2015classification, Jiang2021}, and team communication \cite{kwak2015linguistic, csengun2022players}, our focus shifts to player flexibility and its change over time---specifically a player's willingness to embrace different roles when teamed with strangers.

To dissect the nuances of flexibility, our work is guided by three research questions: \textit{(a)} How is user flexibility operationalized and measured? \textit{(b)} What relationship does flexibility share with a user’s skill level and play style preferences? And crucially, \textit{(c)} which attributes can predict a player's flexibility in both general and temporal contexts?

These questions not only build on foundational player classification and avatar representation research \cite{Williams2010, Jiang2021} but also seek to apply Billieux et al.'s \cite{Billieux2013} findings on the relationship of personal motives driving in-game decisions to ad hoc and ephemeral teams in competitive environments.
Through this lens, we take advantage of the complex nature of player flexibility to shed light on the critical role it plays in the ever-evolving landscape of massive multiplayer team-based games.

\section{Methodology}
Our study leverages LoL's intricate ecosystem, utilizing its diverse competitive modes, developer-defined play styles, and readily available structured data. During a game, players choose one of 168 (and counting) unique \textit{champions}, each with distinct strengths and weaknesses. 
We primarily focus on the most competitive mode, ``Solo Queue'', but also consider the casually competitive ``Flex Queue'' and casual game modes. 
``Solo Queue'' only allows pairs of closely ranked users to queue together, with pairs being disallowed at high ranks. ``Flex Queue'' allows for a broader range of skill levels, with looser rules on team formation. 
These two queues are ``ranked'', and the public and competitive nature of a user's rank sets the stage for our analysis of flexibility in competitive contexts.

\subsection{Dataset}

To construct our dataset, we collected LoL data through Riot Games' public API, with our random sample of users sourced from third-party statistics tracker OP.GG. Players were categorized as either \textit{elite players}—within the top three competitive ranks— or \textit{non-elite players}, with the latter forming the majority of the player base. 
Since the top tier constitutes less than 1\% of the player base, to balance our dataset, we adjust the number of random seeds we select from each rank category to ensure a balanced representative sample of both elite and non-elite players, as detailed in Table~\ref{tab:seed_users}. 
Data collection spanned from mid-June 2023 through the end of September 2023, focusing on users playing in the North American servers. The Riot API allows us to collect up to a user's most recent 100 games played---for active users, this is likely not their entire game history. \footnote{All data collected through Riot's API complies with their terms of service, and is publicly available. We report all findings in aggregate.}
Our dataset contains 4,018 elite users, and 5,621 non-elite users.

We subset this data for different queues (Solo, Flex, All), requiring a minimum of 100 games in that game mode. We find the number of users in our queue-specific datasets decreases as the competitiveness of the game mode increases; the retention of elite players is significantly more than the retention of non-elite players in in more competitive queues. This results in a Solo Queue dataset comprised of $3,119$ elite and $2,304$ non-elite players.

\subsection{Flexibility Score}
Riot Games defines six play styles: \textit{Assassins}, \textit{Fighters}, \textit{Mages}, \textit{Marksman}, \textit{Support} and \textit{Tanks}. 
Each champion has a primary play style and an optional secondary play style. This results in 15 play style pairs (order agnostic) and 6 exclusive play styles for champion categorization. 
We refer to these 21 play styles as a champion's \textit{functional play style}. 

\begin{table}[t]
    \centering \footnotesize
    \begin{tabular}{c|c|c}
         Skill & \textbf{Rank Name}& \textbf{\% Users Included as Seeds} \\
         \hline
         \textbf{Elite}&  Challenger + Grandmaster & 100\%\\
         \textbf{Elite}&  Master & 50\%\\
         \textbf{Non-Elite}& Iron - Diamond& 0.5\%\\
    \end{tabular}
    \caption{Percentage of ladder users retained as seed users. Skill refers to skill classification of elite or non-elite users.}
    \label{tab:seed_users}
\end{table}

A user's \textit{flexibility score}, $F$ is calculated in several steps outlined in Equations~\ref{eq:pi},~\ref{eq:pmax}, and~\ref{eq:flex}: 
\begin{equation}
    P_{i} = \frac{N_i}{N_{games}}
    \label{eq:pi}
\end{equation}
\begin{equation}
    P_{max} = max(P_1,P_2...P_{21})
    \label{eq:pmax}
\end{equation}
\begin{equation}
    F = 1 - P_{norm} \mbox{\quad with \quad } P_{norm}=\frac{1}{21}\sum^{21}_{i=1}|P_{max}-P_i|
    \label{eq:flex}
\end{equation}

In Equation~\ref{eq:pi}, $N_i$ is the number of games that a user has played as a champion falling under \textit{functional play style} category $i$, $N_{games}$ is the total number of games a user has played, and $P_i$ is the percentage of all games that a user has played in that \textit{functional play style}. 
This is done for each of the 21 possible \textit{functional play styles}. 

In Equation~\ref{eq:pmax}, we identify $P_{max}$, or the percentage of a user's most played \textit{functional play style}. Then, we find the average difference between the percentage a user plays each \textit{functional play style} and their most played \textit{functional play style}. 
This average percentage will be larger if a user tends to play more of one specific role, and smaller if a user plays a more even distribution of roles. 

As for Equation~\ref{eq:flex}: to make the \textit{flexibility score}, $F$ more intuitive, we reverse the measure as $1-P_{norm}$, such that an increase in \textit{flexibility score} corresponds to an increase in user flexibility, while a decrease in \textit{flexibility score} results from a user who gravitates towards a narrow set of play styles. 

Finally, we also determine a \textit{play style preference} based on the \textit{functional play style} a user plays the most. Any ties are broken at random. 

\subsection{Temporal Flexibility}
To assess how a player's adaptability evolves, we examine the variation in their \textit{flexibility scores} over time. We apply a rolling window of 10 games over a user's chronological game history, and calculate sequential \textit{flexibility scores} over each window of 10 games. We apply a linear regression model to these rolling scores, yielding a \textit{temporal flexibility coefficient} (Temp Flex Coef). This coefficient indicates the direction and magnitude of change in a player's game style adaptability over time, offering insights into whether they are becoming more versatile or are specializing in certain play styles.

\section{Results}
Despite differences in rank or game mode, a consistent set of top five functional play styles emerged as popular choices among players. 
This pattern suggests there exists a core set of strategies or roles that are widely favored across all players and modes, regardless of the competitiveness or casualness of play.

\begin{figure}[th]
\vspace{1cm} 
\centering
\includegraphics[width=\columnwidth,clip,trim=5 30 15 40]{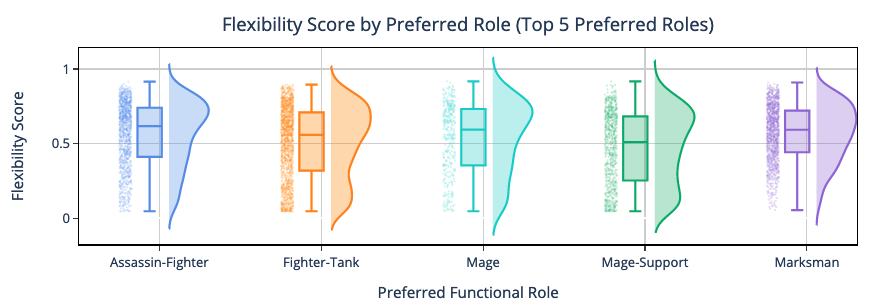}
\caption{Flexibility scores stratified by role preference.}
\label{fig:flex_role}
\end{figure}

\begin{figure*}[th]
\centering
\includegraphics[width=\textwidth,clip,trim=5 40 20 20]{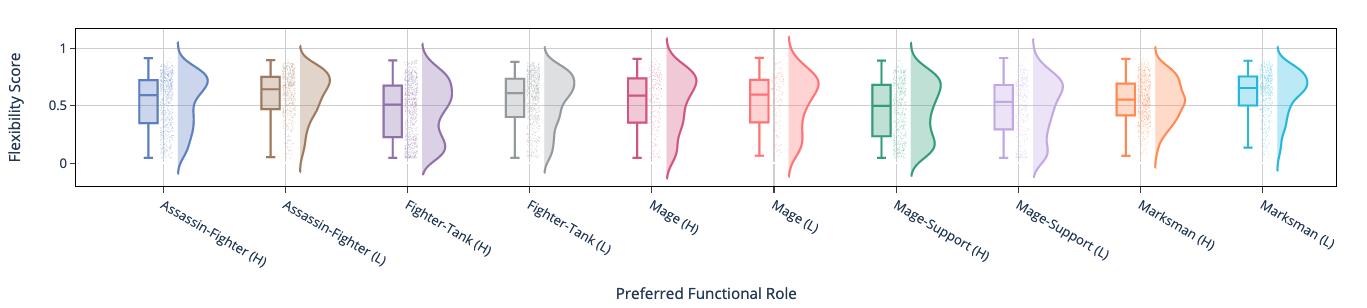}
\caption{Flexibility scores stratified by both role preference and skill classification, where H denotes elite players, and L denotes non-elite players.}
\label{fig:flex_role_skill}
\end{figure*}

\subsection{Flexibility Scores} \label{sec:results_flexscores}
We calculate the general flexibility scores of each player in our dataset for their Solo Queue, Flex Queue and complete match history.  
Our analysis finds that the flexibility scores of elite players are consistently lower than non-elite players' scores.
As queues increase in competitiveness, flexibility scores drop for all users regardless of their rank. 
Our future discussions will concentrate primarily on individuals who have played a minimum of 100 Solo Queue games and their flexibility within these matches.

We explore the impact of functional play style on flexibility by examining the five most popular play styles across all users in our dataset: \textit{Assassin-Fighter}, \textit{Mage-Support}, \textit{Marksman}, \textit{Fighter-Tank}, and \textit{Mage} champions. 
Analyzing the distribution density of flexibility scores by preferred functional role, as depicted in Figure~\ref{fig:flex_role}, reveals that players favoring \textit{Assassin-Fighter}, \textit{Marksman}, and \textit{Mage} roles tend to exhibit greater flexibility. 
Conversely, those preferring \textit{Mage-Support} and \textit{Fighter-Tank} roles demonstrate a bimodal distribution in flexibility scores, with one cluster showing lower flexibility and another, more pronounced cluster, expressing higher flexibility.

Stratification of participants into elite and non-elite categories shows distinct differences in the density of flexibility score distributions between these groups, as shown in Figure~\ref{fig:flex_role_skill}. 
Notably, elite players exhibit a more even distribution across the flexibility score range relative to their non-elite counterparts. 
This discrepancy is particularly evident in the bimodal nature of \textit{Fighter-Tanks} and \textit{Mage-Supports}, where the distribution among elite players contributes to the emergence of distinct higher and lower flexibility clusters. 
This pattern suggests a notable prevalence of ``One Trick Ponies'' players specializing in a single role with limited flexibility—in higher ranks, especially within the \textit{Mage-Support} and \textit{Fighter-Tank} categories.

Interestingly, elite players who favor \textit{Marksman} roles tend to have a more centralized distribution of flexibility scores, contrasting with non-elite \textit{Marksman} players who exhibit more willingness to engage with different play styles.

\subsection{Temporal Flexibility}
Our analysis continues to focus on users that play Solo Queue games in our dataset. 
We calculate each user's temporal flexibility coefficients; as flexibility scores are bound between 0 and 1, the regression coefficients are inherently small. 
Therefore we leverage comparative magnitude and directional trends, as inferred from the value and sign of the coefficient. 

The temporal flexibility coefficients across all players reflects a general increase in flexibility over time. 
This observation, supported by recent findings on the effects of game updates \cite{claypool2017impact, he2021heterogeneous}, is likely influenced by Riot's consistent release of new champions, broadening the spectrum of available play styles and encouraging players to explore novel champion roles. 
Stratifying these coefficients by player skill level uncovers distinct trends: elite players demonstrate an increase in flexibility over time, whereas non-elite players experience a decline. 
More notably, the rate of increase in flexibility among high-ranked players is over sixfold faster than the decrease observed in their low-ranked counterparts. 
This indicates that elite players are more likely to experiment with new play styles, at the expense of competitive success, compared to non-elite players who are less likely to do the same. 

We then consider both a user's preferred functional play style and their rank. 
This analysis finds that elite players favoring \textit{Assassin-Fighter}, \textit{Mage-Support}, and \textit{Marksman} roles show an upward trend in flexibility, while their non-elite peers tend to see a decrease.
Conversely, elite and non-elite players preferring \textit{Fighter-Tank} and \textit{Mage} roles both experience an increase in flexibility, with elite individuals doing so at a significantly faster pace. 
These differences and similarities in temporal flexibility across various play styles and ranks echoes the distinctions in flexibility score distributions discussed in Section~\ref{sec:results_flexscores}, highlighting the complex interplay between player rank, preferred roles, and evolving game dynamics.

\subsection{Predicting Flexibility}
Our final objective is to identify factors that can predict a user's 1) overall flexibility scores and 2) changes to their flexibility scores over time.

\subsubsection{Solo Queue Win Rates}
Initially, we employ multivariate linear regression (MLR) to estimate Solo Queue win rates as a basic validation step. 
This model incorporates variables such as a user's skill level, their overall flexibility score, and the trend in their flexibility score over time (temporal flexibility coefficient). 
We include the dummy variable \textit{non-elite} in order to account for a player's skill classification as an elite or non-elite player. 
As Table~\ref{tab:mvlr_soloq_wr} shows, both skill classification and general flexibility scores inversely correlate with Solo Queue win rates. 
This finding suggests that competitive success in League of Legends is more closely associated with mastery of a limited set of functional play styles (less flexibility) rather than a broad knowledge of many play styles. 
Specifically, lower flexibility scores correlate with higher Solo Queue win rates, reinforcing our previous observations. 
This also aligns with our expectations, as we anticipate that non-elite players will generally have comparatively lower win rates. 
The correlation between the temporal flexibility coefficient and win rate is negative but not statistically significant ($p=0.076$).

\begin{table}[t]
\centering
\begin{threeparttable}
\small
\begin{tabular}{lcc}
\toprule
 & \multicolumn{2}{c}{Model 1}  \\
 & Coefficient & Std. Error  \\
\midrule
Intercept & 0.5128\(^{***}\) &  0.001\\
Non-Elite &  -0.0082\(^{***}\) & 0.001 \\
Flex Score & -0.0096\(^{***}\) & 0.002 \\
Temp Flex Coef & -0.7122 & 0.402 \\
\bottomrule
\end{tabular}
\begin{tablenotes}\footnotesize
\item \textit{Note:} \(*** p < 0.001\)
\end{tablenotes}
\end{threeparttable}
\vspace{0.25cm}
\caption{MLR for Solo Queue WR}
\label{tab:mvlr_soloq_wr}
\end{table}

\subsubsection{Flexibility Scores}
When predicting flexibility scores, we use two distinct multivariate linear regression models. Both of these models include a dummy variable for skill classification (\textit{Non-Elite}) and Solo Queue win rate. 
The first model focuses on temporal trends by including the temporal flexibility coefficient. 
The second model evaluates the impact of competitive engagement on flexibility by using a binary variable, \textit{more ranked}, which indicates if a user plays more Solo Queue games over other modes. 

\begin{table}[t]
\centering
\begin{threeparttable}

 \footnotesize
 \begin{tabular}{@{}lcccc@{}}
\toprule
 & \multicolumn{2}{c}{Model 1} & \multicolumn{2}{c}{Model 2} \\
 & Coefficient & Std. Error & Coefficient & Std. Error \\
\midrule
Intercept & 0.671\(^{***}\) & 0.040 & 0.661\(^{***}\) & 0.040\\
Non-Elite & 0.047\(^{***}\) & 0.006 & 0.048\(^{***}\) & 0.006 \\
Solo Queue WR & -0.313\(^{***}\) & 0.077 & -0.309 & 0.077 \\
Temp Flex Coef & -2.242 & 2.292 &  &  \\
More Ranked & & & 0.0099 & 0.007 \\
\bottomrule
\end{tabular}
\begin{tablenotes}\footnotesize
\item \textit{Note:} \(*** p < 0.001\)
\end{tablenotes}
\vspace{0.25cm}

\caption{MLR for Flexibility Score}
\label{tab:mvlr_flex_score}
\end{threeparttable}
\end{table}

Both models in Table~\ref{tab:mvlr_flex_score} indicate that the temporal flexibility coefficient and the proportion of ranked to casual games are not significant predictors of a user's flexibility score. 
These analyses reaffirm the negative correlation between win rates and flexibility scores as the models suggest that non-elite players tend to have higher flexibility scores. 
The inclusion of a user's preferred play style in a third model (not presented in the tables) identified certain roles as significant predictors of flexibility scores. 
However, this finding should be approached with caution due to uneven distributions of champions across play styles, potentially causing players to adopt other styles out of necessity.

\subsubsection{Temporal Flexibility Coefficient}
Finally, we use another multivariate linear regression model to predict users' temporal flexibility coefficients, considering their skill classification, Solo Queue win rate, general flexibility score, and whether they played more ranked or casual games. 
Surprisingly, only the user's skill classification proves statistically significant in explaining the temporal flexibility coefficient, as detailed in Table~\ref{tab:mvlr_temp_flex_coef}. 
This result implies that non-elite players experience a slight decrease in flexibility over time. 
The lack of statistical significance for other variables might suggest that players are generally set in their flexibility preferences and are resistant to change. 
There is also a possibility that behavioral changes in flexibility require longer periods of time to manifest, as the dataset used for this work covers partial historical data (up to 100 prior games from when our collection was initiated) in addition to four months of real-time collection.

\begin{table}[t]
\centering
\begin{threeparttable}
\small
\begin{tabular}{lcc}
\toprule
 & \multicolumn{2}{c}{Model 1}  \\
 & Coefficient & Std. Error  \\
\midrule
Intercept & 0.0005\(^{*}\) & 0.000 \\
Non-Elite & -0.0001\(^{**}\) & 3.67$e^{-5}$ \\
Solo Queue WR & -0.0008 & 0.000 \\
More Ranked & -5.803$e^{-6}$ & 4.4$e^{-5}$ \\
Flex Score & -7.85$e^{-5}$ & 8.05$e^{-5}$ \\
\bottomrule
\end{tabular}
\begin{tablenotes}\footnotesize
\item \textit{Note:} \(* p < 0.05, ** p < 0.01\)
\end{tablenotes}

\end{threeparttable}
\vspace{0.25cm}
\caption{MLR for Temporal Flex Coef.}
\label{tab:mvlr_temp_flex_coef}
\end{table}

\section{Conclusions}

Our research into flexibility dynamics in League of Legends yields insightful revelations on how player skill, play style preference, and the game's evolving landscape influences a user's overall flexibility and temporal flexibility. 
These insights significantly enhance our understanding of strategic game play patterns and player adaptation, underscoring the nuanced relationship between player behavior and game mechanics.

While our study faced some limitations including rate limited data access, it nonetheless provides an intriguing set of insights in this first-attempt at modeling player flexibility. 

This study marks an important step towards highlighting the significance of play style flexibility when examining player behavior. 
We find that increasing game competitiveness leads to reduced flexibility. 
Users' rank plays a role in predicting their flexibility, and elite players tend to specialize more than non-elite players do. 
This specialization is also correlated with increases in game-based success; however, elite players are often more open to trying new play styles despite this being correlated with a decrease in competitive or game success. 
Flexibility distributions also varies by play style preference and skill classification, as roles with a predominant bimodal flexibility distributions are primarily driven by elite player flexibility distributions. 

In future studies, we hope to extend this work to cover a larger longitudinal dataset and expand to include users from other regions outside of North America, as behavioral tendencies can vary globally. 
We anticipate that this and future work pursing this line of research will contribute to the games and esports landscape, with insights helping game companies cater towards diverse user bases based on game objectives and skill levels. 
However, we also hope that our work will provide insights into the role of flexibility in environments characterized by different degrees of stress and competition, so that we can guide both strategic planning and policy implementation in these dynamic social environments aiming to enhance flexibility or diversity of choice. 

\bibliographystyle{ACM-Reference-Format}
\bibliography{sample-base}


\begin{thebibliography}{11}


\ifx \showCODEN    \undefined \def \showCODEN     #1{\unskip}     \fi
\ifx \showDOI      \undefined \def \showDOI       #1{#1}\fi
\ifx \showISBNx    \undefined \def \showISBNx     #1{\unskip}     \fi
\ifx \showISBNxiii \undefined \def \showISBNxiii  #1{\unskip}     \fi
\ifx \showISSN     \undefined \def \showISSN      #1{\unskip}     \fi
\ifx \showLCCN     \undefined \def \showLCCN      #1{\unskip}     \fi
\ifx \shownote     \undefined \def \shownote      #1{#1}          \fi
\ifx \showarticletitle \undefined \def \showarticletitle #1{#1}   \fi
\ifx \showURL      \undefined \def \showURL       {\relax}        \fi
\providecommand\bibfield[2]{#2}
\providecommand\bibinfo[2]{#2}
\providecommand\natexlab[1]{#1}
\providecommand\showeprint[2][]{arXiv:#2}

\bibitem[Billieux et~al\mbox{.}(2013)]%
        {Billieux2013}
\bibfield{author}{\bibinfo{person}{Jo{\"e}l Billieux}, \bibinfo{person}{Martial Van~der Linden}, \bibinfo{person}{Sophia Achab}, \bibinfo{person}{Yasser Khazaal}, \bibinfo{person}{Laura Paraskevopoulos}, \bibinfo{person}{Daniele Zullino}, {and} \bibinfo{person}{Gabriel Thorens}.} \bibinfo{year}{2013}\natexlab{}.
\newblock \showarticletitle{Why do you play World of Warcraft? An in-depth exploration of self-reported motivations to play online and in-game behaviours in the virtual world of Azeroth}.
\newblock \bibinfo{journal}{\emph{Computers in Human Behavior}} \bibinfo{volume}{29}, \bibinfo{number}{1} (\bibinfo{year}{2013}), \bibinfo{pages}{103--109}.
\newblock


\bibitem[Claypool et~al\mbox{.}(2015)]%
        {claypool2015surrender}
\bibfield{author}{\bibinfo{person}{Mark Claypool}, \bibinfo{person}{Jonathan Decelle}, \bibinfo{person}{Gabriel Hall}, {and} \bibinfo{person}{Lindsay O'Donnell}.} \bibinfo{year}{2015}\natexlab{}.
\newblock \showarticletitle{Surrender at 20? Matchmaking in league of legends}. In \bibinfo{booktitle}{\emph{2015 IEEE Games Entertainment Media Conference (GEM)}}. IEEE, \bibinfo{pages}{1--4}.
\newblock


\bibitem[Claypool et~al\mbox{.}(2017)]%
        {claypool2017impact}
\bibfield{author}{\bibinfo{person}{Mark Claypool}, \bibinfo{person}{Artian Kica}, \bibinfo{person}{Andrew La~Manna}, \bibinfo{person}{Lindsay O’Donnell}, {and} \bibinfo{person}{Tom Paolillo}.} \bibinfo{year}{2017}\natexlab{}.
\newblock \showarticletitle{On the impact of software patching on gameplay for the league of legends computer game}.
\newblock \bibinfo{journal}{\emph{The Computer Games Journal}}  \bibinfo{volume}{6} (\bibinfo{year}{2017}), \bibinfo{pages}{33--61}.
\newblock


\bibitem[Eggert et~al\mbox{.}(2015)]%
        {eggert2015classification}
\bibfield{author}{\bibinfo{person}{Christoph Eggert}, \bibinfo{person}{Marc Herrlich}, \bibinfo{person}{Jan Smeddinck}, {and} \bibinfo{person}{Rainer Malaka}.} \bibinfo{year}{2015}\natexlab{}.
\newblock \showarticletitle{Classification of player roles in the team-based multi-player game dota 2}. In \bibinfo{booktitle}{\emph{Entertainment Computing-ICEC 2015: 14th International Conference, ICEC 2015, Trondheim, Norway, September 29-Ocotober 2, 2015, Proceedings 14}}. Springer, \bibinfo{pages}{112--125}.
\newblock


\bibitem[He et~al\mbox{.}(2021)]%
        {he2021heterogeneous}
\bibfield{author}{\bibinfo{person}{Yuzi He}, \bibinfo{person}{Christopher Tran}, \bibinfo{person}{Julie Jiang}, \bibinfo{person}{Keith Burghardt}, \bibinfo{person}{Emilio Ferrara}, \bibinfo{person}{Elena Zheleva}, {and} \bibinfo{person}{Kristina Lerman}.} \bibinfo{year}{2021}\natexlab{}.
\newblock \showarticletitle{Heterogeneous effects of software patches in a multiplayer online battle arena game}. In \bibinfo{booktitle}{\emph{Proceedings of the 16th International Conference on the Foundations of Digital Games}}. \bibinfo{pages}{1--9}.
\newblock


\bibitem[Jiang et~al\mbox{.}(2021)]%
        {Jiang2021}
\bibfield{author}{\bibinfo{person}{Julie Jiang}, \bibinfo{person}{Danaja Maldeniya}, \bibinfo{person}{Kristina Lerman}, {and} \bibinfo{person}{Emilio Ferrara}.} \bibinfo{year}{2021}\natexlab{}.
\newblock \showarticletitle{The wide, the deep, and the maverick: Types of players in team-based online games}.
\newblock \bibinfo{journal}{\emph{Proceedings of the ACM on Human-Computer Interaction}} \bibinfo{volume}{5}, \bibinfo{number}{CSCW1} (\bibinfo{year}{2021}), \bibinfo{pages}{1--26}.
\newblock


\bibitem[Kim et~al\mbox{.}(2017)]%
        {kim2017makes}
\bibfield{author}{\bibinfo{person}{Young~Ji Kim}, \bibinfo{person}{David Engel}, \bibinfo{person}{Anita~Williams Woolley}, \bibinfo{person}{Jeffrey Yu-Ting Lin}, \bibinfo{person}{Naomi McArthur}, {and} \bibinfo{person}{Thomas~W Malone}.} \bibinfo{year}{2017}\natexlab{}.
\newblock \showarticletitle{What makes a strong team? Using collective intelligence to predict team performance in League of Legends}. In \bibinfo{booktitle}{\emph{Proceedings of the 2017 ACM conference on computer supported cooperative work and social computing}}. \bibinfo{pages}{2316--2329}.
\newblock


\bibitem[Kwak and Blackburn(2015)]%
        {kwak2015linguistic}
\bibfield{author}{\bibinfo{person}{Haewoon Kwak} {and} \bibinfo{person}{Jeremy Blackburn}.} \bibinfo{year}{2015}\natexlab{}.
\newblock \showarticletitle{Linguistic analysis of toxic behavior in an online video game}. In \bibinfo{booktitle}{\emph{Social Informatics: SocInfo 2014 International Workshops, Barcelona, Spain, November 11, 2014, Revised Selected Papers 6}}. Springer, \bibinfo{pages}{209--217}.
\newblock


\bibitem[Sapienza et~al\mbox{.}(2018)]%
        {sapienza2018individual}
\bibfield{author}{\bibinfo{person}{Anna Sapienza}, \bibinfo{person}{Yilei Zeng}, \bibinfo{person}{Alessandro Bessi}, \bibinfo{person}{Kristina Lerman}, {and} \bibinfo{person}{Emilio Ferrara}.} \bibinfo{year}{2018}\natexlab{}.
\newblock \showarticletitle{Individual performance in team-based online games}.
\newblock \bibinfo{journal}{\emph{Royal Society open science}} \bibinfo{volume}{5}, \bibinfo{number}{6} (\bibinfo{year}{2018}), \bibinfo{pages}{180329}.
\newblock


\bibitem[{\c{S}}eng{\"u}n et~al\mbox{.}(2022)]%
        {csengun2022players}
\bibfield{author}{\bibinfo{person}{Sercan {\c{S}}eng{\"u}n}, \bibinfo{person}{Joao~M Santos}, \bibinfo{person}{Joni Salminen}, \bibinfo{person}{Soon-gyo Jung}, {and} \bibinfo{person}{Bernard~J Jansen}.} \bibinfo{year}{2022}\natexlab{}.
\newblock \showarticletitle{Do players communicate differently depending on the champion played? Exploring the Proteus effect in League of Legends}.
\newblock \bibinfo{journal}{\emph{Technological Forecasting and Social Change}}  \bibinfo{volume}{177} (\bibinfo{year}{2022}), \bibinfo{pages}{121556}.
\newblock


\bibitem[Williams et~al\mbox{.}(2011)]%
        {Williams2010}
\bibfield{author}{\bibinfo{person}{Dmitri Williams}, \bibinfo{person}{Tracy~LM Kennedy}, {and} \bibinfo{person}{Robert~J Moore}.} \bibinfo{year}{2011}\natexlab{}.
\newblock \showarticletitle{Behind the avatar: The patterns, practices, and functions of role playing in MMOs}.
\newblock \bibinfo{journal}{\emph{Games and culture}} \bibinfo{volume}{6}, \bibinfo{number}{2} (\bibinfo{year}{2011}), \bibinfo{pages}{171--200}.
\newblock


\end{thebibliography}

\end{document}